\documentclass[preprint,pra,aps,amsmath,amssymb,showpacs,superscriptaddress]{revtex4-1}

\usepackage{graphics}
\usepackage{epsfig}
\usepackage{hyperref}
\usepackage{amsfonts}
\usepackage{amssymb}
\usepackage{amsmath}
\usepackage{gensymb}

\usepackage{graphicx}
\usepackage{dcolumn}
\usepackage{bm}

\newcommand{\be}{\begin{equation}}
\newcommand{\ee}{\end{equation}}

\begin{document}
\title{Evidence for magnetic crystallization waves at the surface of $^3$He crystal}
\author{I.\,Todoshchenko}
\affiliation{Low Temperature Laboratory, Department of Applied Physics, Aalto University, P.O. Box 15100, FI-00076 Espoo, Finland}

\author{A.\,Savin}
\affiliation{Low Temperature Laboratory, Department of Applied Physics, Aalto University, P.O. Box 15100, FI-00076 Espoo, Finland}
\affiliation{OtaNano, Aalto University, School of Science, P.O. Box 15100, FI-00076 Aalto, Finland}

\author{P.J.\,Hakonen}
\affiliation{Low Temperature Laboratory, Department of Applied Physics, Aalto University, P.O. Box 15100, FI-00076 Espoo, Finland}
\affiliation{QTF Centre of Excellence, Department of Applied Physics, Aalto University, P.O. Box 15100, FI-00076 Aalto, Finland}

\date{February 2024}
\begin{abstract}
Ultralow temperature crystals of the helium isotopes $^3$He and $^4$He are intriguing quantum systems. Deciphering the complex features of these unusual materials has been made possible in large part by Alexander Andreev's groundbreaking research. In 1978, Andreev and Alexander Parshin predicted the existence of melting/freezing waves at the surface of a solid $^4$He crystal, which was subsequently promptly detected. Successively, for the fermionic $^3$He superfluid/solid interface, even more intricate crystallization waves were anticipated, although they have not been observed experimentally so far. In this work, we provide preliminary results on $^3$He crystals at the temperature $T = 0.41$\;mK, supporting the existence of spin supercurrents in the melting/freezing waves on the crystal surface below the antiferromagnetic ordering temperature $T_N= 0.93$\;mK, as predicted by Andreev. The spin currents that accompany such a melting-freezing wave make it a unique object, in which the inertial mass is distinctly different from the gravitational mass.  
\end{abstract}

\maketitle

\section{Introduction}

The emergence of crystallization waves at the liquid-solid interface of quantum crystals represents an intriguing notion which, for the most part, arose from the insightful theoretical work of Alexander Andreev \cite{Andreev1978,Moscow,Andreev1982,Andreev1993}. Besides $^4$He crystals, Andreev applied these fundamental quantum mechanical concepts to show that melting/freezing waves are, indeed, expected to emerge at the surface of $^3$He crystals at sub-milliKelvin temperatures \cite{Andreev1993,Andreev1995}. In particular, the presence of a magnetic field leads to stronger polarization of the solid compared with the fluid \cite{Chapelier}, which results in significant spin currents near the moving interface. Below the Neel temperature $T_N \simeq 1$\;mK the solid forms an antiferromagnetically ordered state, which influences the temperature dependence of the entropy of the solid phase significantly. While cooling further to lower temperatures, once the entropy difference between liquid and solid phases has been substantially reduced, and the critical speeds for magnons in the solid or for pair breaking in the superfluid are not exceeded, high mobility for the superfluid/solid interface can be reached. This large mobility facilitates the macroscopic melting/freezing waves at the interface.

According to Andreev, spin currents accompanying the moving $^3$He liquid-solid interface in the presence of a magnetic field add to the wave's inertia since their energy is proportional to the square of the interface's velocity. This leads to a violation of Einstein's fundamental {\it Equivalence principle} which states that inertial mass is equivalent to gravitational mass \cite{Einstein}. Provided that the magnetic contribution to the energy spectrum of the waves is observed, this would give a unique example of a particle, a melting-freezing ripplon, whose inertial mass is not equal to gravitational mass.

The antiferromagnetic structure of solid $^3$He at $T < T_N$ is formed of the so-called up-up-down-down (uudd) state \cite{OsheroffReview}, provided that the magnetic field $H < H_N \sim T_N/\mu$, where $\mu$ denotes the magnetic moment of a $^3$He atom. Under such field strengths, while being below $T_N$, the B-phase provides the stable superfluid phase \cite{VW}. The spin polarization involved in the crystallization waves is dependent on the magnetic susceptibility of the liquid and solid phases, $\chi_L$ and $\chi_S$, respectively. With the antiferromagnetic vector of the uudd structure oriented perpendicular to $H$, the solid susceptibility $\chi_S \gg \chi_L$, which leads to a clear difference of spin polarization between the phases and thereby to the presence of spin currents in the crystallization waves in this system. 

The spin dynamics of ordered $^3$He phases is governed by their characteristic NMR frequencies, which are closely comparable in the superfluid B-phase and in the uudd crystal, $f_L \sim 3 \times 10^5$ Hz \cite{Ahonen_NMR,Hakonen_NMR} and $f_S \simeq 8 \times 10^5$\,Hz \cite{Osheroff1980}, respectively. The spin wave velocity in the solid phase amounts to $u_S=7.7$\;cm/s \cite{Osheroff1980b,Busch1987,Ni1994} while, in superfluid $^3$He-B, $u_L\simeq 8$\;m/s was found by Osheroff {\it et al.}~\cite{Osheroff1977}.  As the spin wave speed in the solid $u_S \ll u_L$, the spin wave velocity in the liquid, the characteristic dipole length scales are quite different: $\ell_S=u_S/f_S \sim 10^{-7}$\,m in contrast to $\ell_L=u_L/f_L \sim 10^{-5}$\,m.

The surface stiffness of the liquid-solid interface can be written as $\tilde \alpha  = \alpha  + \frac{{{\partial ^2}\alpha }}{{\partial {\varphi ^2}}}$ where the surface energy $\alpha(\varphi)$ is expanded in terms of variable $\varphi$ denoting the orientation of the surface (see e.g.~Refs.~\onlinecite{Nozieres1992,Balibar2005}). Neglecting gravity terms, the potential energy for the melting freezing waves with wave vector $q$ is given by the effective surface stiffness $\tilde \alpha q^2$. Therefore, we may write for the dispersion relation of the oscillations
\be
{\omega ^2}(q) = \frac{{\tilde \alpha {q^2}}}{{M(q)}},
\ee
where the total mass $M(q) = M_m + M_s$ is a sum of the mass in the kinetic energy, $M_m$, and the mass of the kinetic energy of the spin supercurrent, $M_s$. The spin contribution to the effective mass becomes dominant already at rather small magnetic fields. The total mass may be written as
\be
M(q) = \rho d\left[ {{{\left( {\frac{H}{H_0}} \right)}^2} + {{\left( {\frac{{\Delta \rho }}{\rho }} \right)}^2}\frac{1}{{qd}}} \right]
\ee
where the first term describes the spin current contribution to the mass, while the latter represents the regular superflow contribution. Here we have introduced a characteristic field ${H_0} = \frac{{{u_L}}}{{{\chi_S}}}\sqrt {\rho {\chi _L}}$ and a length scale
\be
d = \frac{{{\chi _L}u_L^2}}{{{\chi _S}{u_S}{f_S} + {\chi _L}{u_L}{f_L}}}=\frac{1}{{\frac{{{\chi _S}}}{{{\chi _L}}}(\frac{{{u_S}}}{{{u_L}}})^2\frac{1}{{{\ell _S}}} + \frac{1}{{{\ell _L}}}}}.
\ee
Using an estimate $\frac{\chi _S}{\chi _L} \sim 200$ and the spin wave velocities quoted above, we obtain $d \sim 4.3$ $\mu$m, which is about two times smaller than the dipole length in $^3$He-B at high pressure, $\xi_D=10$\;$\mu{m}$ \cite{Hakonen_NMR}. Furthermore, by taking 25.5 cm$^3$/mole for the molar volume and  $\chi_L = 1.1 \times 10^{-7}$ for the liquid susceptibility \cite{WheatleyRMP}, we obtain $H_0= 0.4$ T for the characteristic field. Because of the size of our experimental resonator (square 5\;mm by 5\;mm), the capillary length becomes relevant, and we need to employ the full form 

\begin{equation}
\label{eq:spectrum}
\left[ {\frac{{{{\left( {{\rho _S} - {\rho _L}} \right)}^2}}}{{{\rho _L}q}} + {\rho _L}{d}\frac{{{H^2}}}{{H_0^2}}} \right]{\omega ^2} = \left( {{\rho _S} - {\rho _L}} \right)g + \gamma {q^2}
\end{equation}

\noindent
for the dynamics of the interface. The resonator size yields for the capillary term in Eq.\;\ref{eq:spectrum} $\gamma{q}^2\sim5$\;g$/$(cm$^2$s), which is approximately equal to the gravitational contribution $({\rho_S} - {\rho_L})g$. 

The mobility of the solid-liquid interface was estimated by Korshunov and Smirnov 
 \cite{KorshunovSmirnov} as $k=\rho_S\hbar^3(u_S/k_BT)^4$, limited by the repulsion between thermal magnons and the interface. The quality factor of the wave according to this prediction is too small ($Q < 1$) at temperatures above $\sim$0.2\;mK \cite{Andreev1995,Andreev1996}. However, our measurements of the growth rates of $^3$He facets have shown that the mobility is underestimated from the above equation for $k$ \cite{step}. We have observed that the elementary steps on facets reach the critical velocity, that is the magnon velocity in solid $u_S=8$\;cm$/$s, at overpressures on the order of $\delta{p}=0.5$\;mbar at $T=0.93$\;mK, which gives the lowest limit estimate for the growth coefficient $k_{min}(0.93 \mathrm{mK})=u_S/\Delta\mu=u_S\rho_S\rho_L/\delta{p}(\rho_S-\rho_L)/=4.5$\;s$/$m \cite{step}, while the estimation of Korshumov and Smirnov \cite{KorshunovSmirnov} gives only 0.2\;s$/$m. The uncertainty in the theoretical estimate, the reason for which is not yet clear, is thus a factor of 20 which would raise the upper limit of the necessary operation temperature range to $\sim 0.4$\;mK for the observation of weakly damped crystallization waves in the $^3$He crystal.

\section{Preliminary experiments in Helsinki}

In our preliminary experiments, we used an experimental cell embedded in the copper demagnetization stage with $\sim100$\;m$^2$ of sintered surface area \cite{dry}. A square resonator for crystallization waves was made using four copper plates placed at 45\degree\ tilting angle with respect to the vertical axis. Such a 45\degree\ angle corresponds approximately to the contact angle of the solid-liquid interface, and this inclination of the walls makes the interface nearly horizontal inside the resonator. On one of the copper plates, a capacitor is wound using two tightly attached copper wires, exactly as in the works on crystallization waves in $^4$He \cite{Moscow,Agnolet}. The interface between the solid and liquid phases is tuned to touch the capacitor and to be slightly below a quartz tuning fork with its prongs oriented downwards towards the solid. To adjust the interface to this position is difficult, because the melting curve of $^3$He has a deep, about 5\;bar, minimum at $T_{min}=317$\;mK, and a solid plug does form in the filling line when cooling below $T_{min}$, which prevents further tuning efforts. 

The filling of the cell was done at a high temperature, typically around 1\;K, at which there is no solid. If the condensed amount of liquid helium is not correct, then the fork might be frozen in with solid after cooling below 1\;mK; alternatively, the fork may remain too far from the solid, which results in a too low sensitivity for the motion of the interface. One can calculate the needed density of pressurized liquid helium $\rho_L$ to obtain the expected relative amount of solid $x$ in the cell at low temperature from the relation $\rho_L(T)=[(1-x)\rho_L(0)+x\rho_S(0)]$. To find the needed pressure at high temperatures, we have used $P-V-T$ relations measured by Grilly \cite{Grilly}. However, the geometry of the cell was quite complex, and the volumes of the filling line and the heat exchangers were not known exactly. That is why we had to make several iterations for the condensed liquid amount at 1\;K. After each filling we cooled down below 1\;mK and checked whether the amount of solid was proper for our measurements, i.e.~whether the interface is on the capacitor, and close enough to the prongs of the fork which still resided fully in the liquid.

\begin{figure}[htb]
\includegraphics[width=0.98\linewidth]{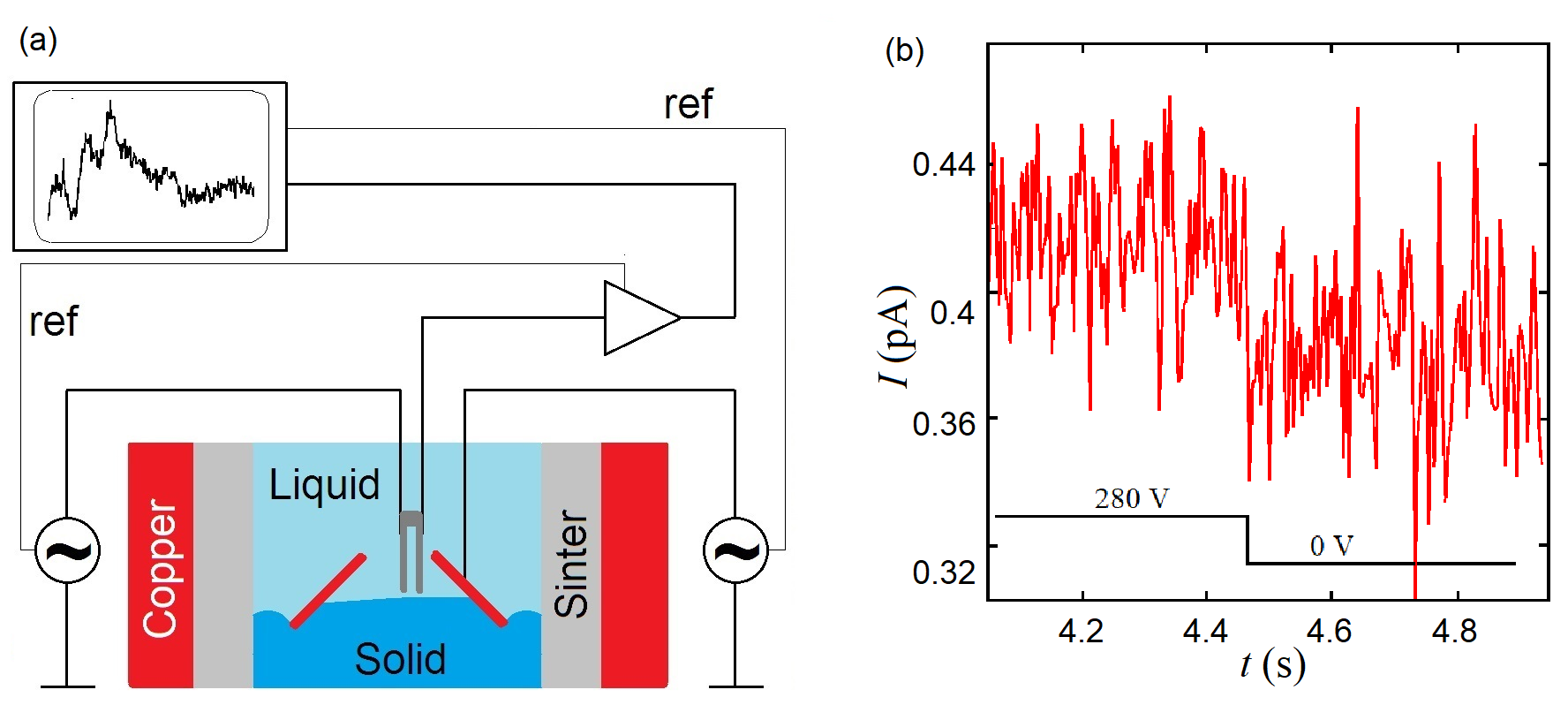}
\caption{\label{fig:meander} Excitation/detection of melting/freezing waves at the solid-liquid interface: a) Schematic experimental setup of our square-shaped resonator (cross-sectional view). The four copper walls delineating the resonator area are tilted by the contact angle of 45\degree\, in order to obtain a nearly horizontal interface. The fork is positioned close to one of the walls to stay away from nodes in the fundamental mode and its even harmonics. The capacitor is wound on the wall that is closest to the fork. b) Height response induced by a voltage step as seen in the Y-quadrature current of the fork oscillating near the resonance. Solid helium is drawn to the strong electric field created by a high voltage supplied to the capacitor, raising the interface level on the capacitor. As a result, the level of solid helium in the center of the resonator is lowered, and the effective mass of liquid helium involved in the oscillations of the fork decreases. Consequently, the resonant frequency of the fork increases proportionally to the interfacial deflection. 
}
\end{figure}

When a high voltage is applied to the capacitor, the interface near the wall is drawn up close to the capacitor, because the solid has a larger electrical susceptibility than the less-dense liquid. An AC voltage applied to the capacitor will therefore induce a crystallization wave if the relaxation rate of the surface is fast enough. If the frequency of the AC voltage squared coincides with the resonance frequency of the cavity, a standing wave will be excited. Estimation of the resonant frequency is simple: the wave vector is obtained from the wavelength set by the dimensions of the cavity (side wall length $L$), i.e. $q=\pi/L$, which is used to evaluate the frequency from Eq.\;\ref{eq:spectrum}. 

\begin{figure}[htb]
\includegraphics[width=0.98\linewidth]{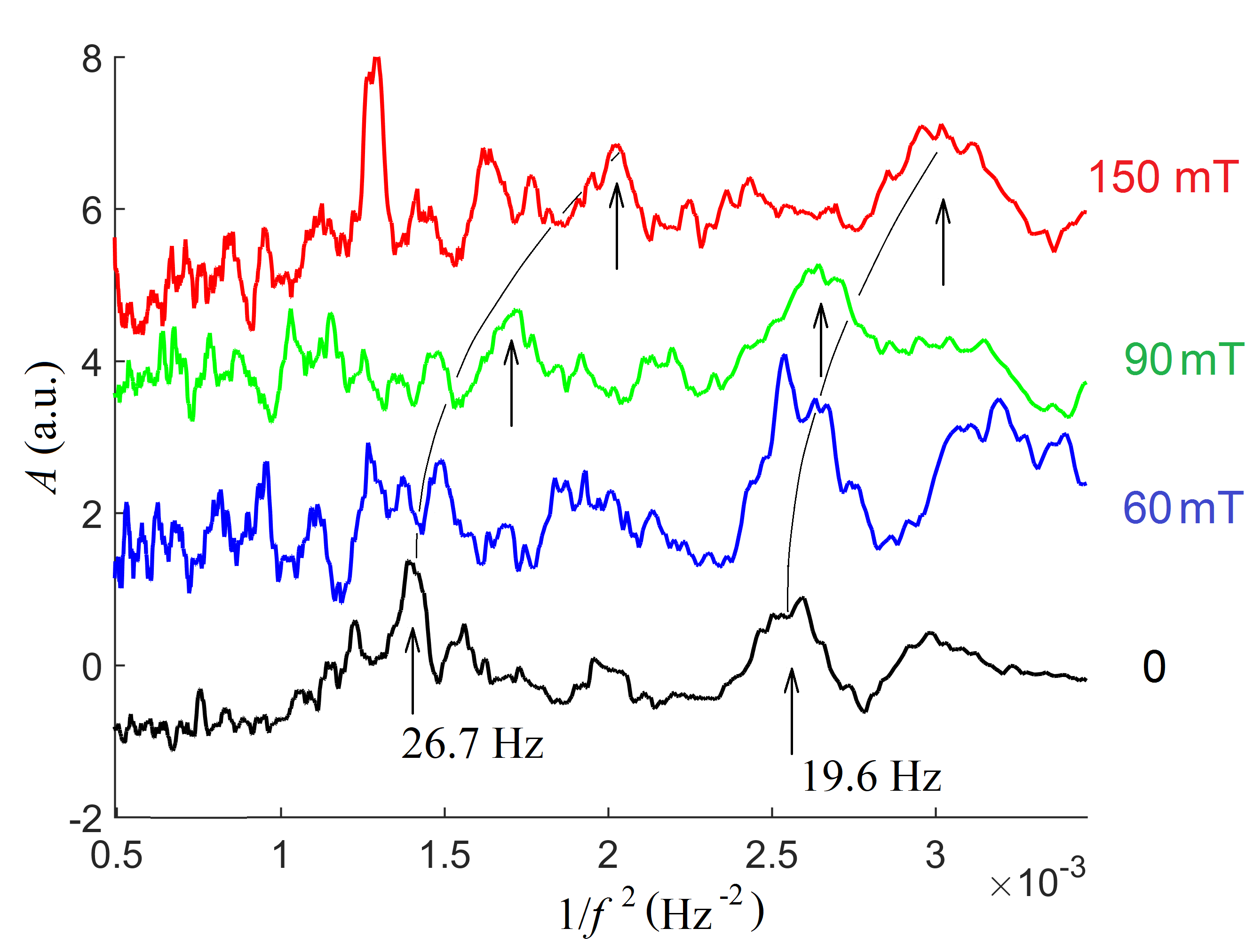}
\caption{\label{fig:waves} Fork current vs. frequency of the excitation voltage applied to the capacitor in the frequency range 17;...\;44\;Hz measured at magnetic fields of $B=0$, 60, 90, and 150 mT (marked in the figure). The arrows indicate our identification for prospective standing waves at the interface. We attribute them to the fundamental mode and its first harmonics. The overlaid thin black lines depict the quadratic magnetic field dependence of the resonances obtained from Eq.\;\ref{eq:spectrum}: $\frac{1}{{{f^2}}} = \frac{1}{g}\left[ {\frac{{\Delta {\rho ^2}}}{{\rho q}} + \frac{{\rho d}}{{\Delta \rho }}{{\left( {\frac{H}{{{H_0}}}} \right)}^2}} \right]$. 
}
\end{figure}

 For the detection of the crystal surface displacement, we employed the oscillating fork scheme tested by our team earlier in the experiments on crystallization waves in $^4$He \cite{4He}. A quartz tuning fork is driven at its resonant frequency where the sensitivity of its out-of-phase signal to a change in the resonator mass is maximal. When the solid-liquid interface near the fork moves, the involved hydrodynamic mass of the liquid helium also changes, which induces a shift of the resonant frequency of the fork. The ensuing change in the out-of-phase current is amplified and demodulated at the double frequency of excitation voltage. The sensitivity of such a double-resonance scheme rapidly increases with decreasing distance between the fork's prongs and the surface of the crystal. The maximum possible sensitivity (when prongs almost touch the surface) is as high as 1\;kHz$/$mm i.e.~1\;mHz/nm \cite{4He}.

In our initial experiments, we determined the relaxation time of the crystal surface. We cooled the solid-liquid interface to the lowest temperature, 0.4\;mK $\approx$ 0.16 $T_c$, and applied a high voltage step change to the capacitor. The out-of-phase signal from the fork due to the shift of resonance reflects the movement of the crystal surface, as shown in Fig.\;\ref{fig:meander}. The time constant of the lock-in amplifier, and the time interval between points were set to $\Delta{t}=3$\;ms. The measured response time was within the time spacing of the points, which yields a conservative estimate  $\tau\lesssim3$\;ms for the relaxation time of the surface of the crystal. Hence, standing crystallization waves with frequencies up to 300\;Hz are possible at these temperatures.

In our measurements we have applied sinusoidal AC high voltage and scan the range 1\;...\;25\;Hz; the fork was staying in its resonance, and the out-of phase signal was demodulated by computer at twice the capacitor frequency. At least two resonant peaks of standing crystallization waves have been observed in this way in zero magnetic field, see Fig.\;\ref{fig:waves}, black curve. With increasing magnetic field these lines move towards lower frequency due to the magnetic inertia. The dependence is quadratic, as far as the poor signal-to-noise ratio allows to determine, as predicted by Andreev. However, the amplitude of the shift is somewhat higher than estimated from Eq.\;\ref{eq:spectrum}. Possible reasons for the discrepancy are discussed in Sects. \ref{Disc},\ref{Magn}.

\section{Discussion} \label{Disc}
Dissipation of surface oscillations via excitations with a linear  spectrum $\epsilon= u \hbar q$ is proportional to $(T/u)^4$. Consequently, excitation of spin waves in the solid phase is the largest dissipation channel for the melting/freezing waves in $^3$He. This dissipation  depends also on the wave vector $q$, and when $qd < 1$ and $H < H_0$, the $Q$ factor can be estimated from
\be \label{eq:quality}
Q \sim A{\left( {\frac{{{T_N}}}{T}} \right)^4}\left[ {\frac{H}{{{H_0}}}qd + \frac{{\Delta \rho }}{\rho }\sqrt {qd} } \right],
\ee
which is valid only to an order of magnitude. The prefactor $A=10^{-2}$ when estimated using the formulation of Korshunov and Smirnov \cite{KorshunovSmirnov}. However, the experimental results of Tsepelin {\it et al.}~ \cite{Tsepelin2002} indicate a smaller step energy than is presumed in the standard theory of Andreev and coworkers. The measured step energy is approximately three orders of magnitude smaller than typically estimated. Consequently, the prefactor $A \sim 10$ is possible in Eq.\;(\ref{eq:quality}), which makes crystallization waves in $^3$He with $qd \sim \pi \cdot 10^{-3}$ realizable up to temperatures of $\sim 0.4$ mK. For Eq. (\ref{eq:quality}), the optimum quality factor is reached around $H\sim H_0$ and $qd \sim 1$. Another possible reason for the observation of crystallization waves up to $ T \sim 0.4$ mK is discussed in Sect. \ref{Magn}.

When the quality factor $Q \sim 1$, there may be a strong reduction of the frequency owing to  dissipation. With decreasing temperature, the $Q$ factor grows, which may lead to a increase in the resonant frequency of oscillations at the crystal interface. Conversely, an increase of $T$ will lead to a reduction of the resonance frequency, which could ultimately bring even higher harmonics to the frequency range of observation. Hence, a high density of magnetic field and temperature values are necessary for reliable analysis. Lacking this large amount of data, we may only tentatively conclude that we see the fundamental crystal oscillation frequency although higher harmonics could yield an increased $Q$ factor. The frequency of these higher harmonics can vary due to changes in the quality factor of the resonances, in qualitative agreement with Eq. \ref{eq:quality}. However, the reduction of the resonance frequency is also in accordance with the increase of the effective mass due to spin supercurrents taking part in the melting/freezing waves. For more definite conclusions, better knowledge of the wave vector magnitude is required in the experiments, for example, by making a microfabricated cell with well controlled small size.

\section{Magnetic surface} \label{Magn}

In his famous papers on magnetic crystallization waves, Andreev made many important predictions \cite{Andreev1996,PLTPh}. The first one was the contribution from spin currents to the inertia of the wave (Eq.\;\ref{eq:spectrum}). This is a very intriguing idea because it proposes a particle (quantum of the surface wave) that has an inertial mass different from the gravitational mass; such a particle has never been identified prior to the present work. Although our measurements agree qualitatively with the prediction by Andreev, namely the quadratic magnetic field dependence of the mass of the waves, the magnitude of this effect is experimentally observed to be much larger than predicted.

This discrepancy may be a sign of the formation of a ferromagnetic layer near the surface of the antiferromagnetic crystal \cite{FM,Meyerovich1981,step}. This effect is similar to the formation of a ferromagnetic polaron in the vicinity of a vacancy in solid $^3$He, described by Andreev \cite{Andreev1976,PLTPh,polaron}. The underlying reason is the enhancement of (ferromagnetic) pair exchange due to the reduction in the density. The ferromagnetically ordered solid layer produces much stronger spin currents on melting/growing, thereby increasing the magnetic mass of the crystallization wave. Moreover, spin currents in the solid may also become significant and contribute to the effect. 

The thickness $h$ of the ferromagnetic layer can be obtained by minimizing free energy per unit area $F=\pi^2\hbar^2/2Mh^2+k_BT(h/a)\log{2}$, $h=(\pi^2a\hbar^2/Mk_BT\log{2})^{1/3} \simeq 4a$ at $T=0.4$\;mK. Here $a$ is the lattice constant and $M$ is the mass of a vacancy taken equal to the atomic mass. In this ferromagnetic layer with reduced density, the exchange energy is intermediate between those of the bulk solid and the liquid. Correspondingly, the magnon velocity $u_{mag}$ in the layer is also intermediate between those of the solid, $u_S=7.7$\;cm$/$s, and of the superfluid, $u_L=8$\;m/s, $u_S<u_{mag}<u_L$. Consequently, the dissipation proportional to $(T/u_{mag})^4$ is greatly reduced, which explains the unexpectedly high quality factor observed for the crystallization waves in these measurements at $ T=0.41$\;mK.

\section{Summary}

We have discussed the observation of crystallization waves in $^3$He on the basis of first experiments performed in a 0.5-cm-sized square resonator. At a temperature of $T = 0.41\;\mathrm{mK}=0.16\;T_c$, two modes of standing crystallization waves were observed by means of a tuning fork detector oscillating in the vicinity of the surface. Both resonance lines shifted significantly with the magnetic field towards lower frequencies, indicating additional inertia due to spin currents accompanying the motion of the surface. The magnetic contribution to the mass has been predicted by Andreev, and our preliminary measurements do support the existence of such a unique wave in which the inertial mass is distinctly different from the gravitational mass.

More experiments are definitely needed to confirm these findings at lower temperatures and higher magnetic fields. Furthermore, experiments using a compressible cell \cite{Hata1989} would be particularly interesting as this would facilitate adjusting the liquid-solid interface position really close to the fork detector.

\section*{Acknowledgements}

This work was supported by the Academy of Finland (AF) projects 341913 (EFT) and by the European Union’s Horizon 2020 Research and Innovation Programme, under Grant Agreement No.~824109 (EMP). The experimental work benefited from the Aalto University OtaNano/LTL infrastructure.

\end{document}